\documentclass[aps,PRB,twocolumn,amsmath,superscriptaddress, showpacs,floatfix,reprint]{revtex4-2}
\usepackage{amsmath, nccmath}
\usepackage{amssymb}
\usepackage{bm}
\usepackage{braket}
\usepackage{natbib}
\usepackage{leftidx}
\usepackage{graphicx}    
\usepackage{verbatim}   
\usepackage{color}      
\usepackage{subfigure}  
\usepackage{hyperref}   
\usepackage{dcolumn}    
\usepackage{textcomp}
\usepackage{float}
\usepackage{threeparttable}
\usepackage{titlecaps}
\usepackage{multirow}
\usepackage{lipsum}
\hypersetup{
	colorlinks=true,
	citecolor=blue,
	filecolor=black,
	linkcolor=blue,
	urlcolor=cyan
}
\usepackage{mathastext}
\usepackage{mathptmx}
\usepackage{enumitem}
\DeclareGraphicsExtensions{.png,.pdf,.tif}
\usepackage{pict2e}
\usepackage[dvipsnames]{xcolor}
\usepackage[normalem]{ulem}

\begin{document}
\title{Anomalous Lattice Effect Originated Metal-Insulator Transition in FeSe$_{\it{x}}$}
\author{Shubham Purwar}
\affiliation{Department of Condensed Matter and Materials Physics, S. N. Bose National Centre for Basic Sciences, Kolkata, West Bengal, India, 700106.}
\author{Shinjini Paul}
\affiliation{Department of Condensed Matter and Materials Physics, S. N. Bose National Centre for Basic Sciences, Kolkata, West Bengal, India, 700106.}
\author{Kritika Vijay}
\affiliation{Accelerator Physics and Synchrotrons Utilization Division, Raja Ramanna Centre for Advanced Technology, Indore 452013, India}
\affiliation{Homi Bhabha National Institute, Training School Complex, Anushakti Nagar, Mumbai, 400094, India}
\author{R. Venkatesh}
\affiliation{UGC-DAE Consortium for Scientific Research, DAVV Campus, Indore 452001, MP, India.}
\author{Soma Banik}
\affiliation{Accelerator Physics and Synchrotrons Utilization Division, Raja Ramanna Centre for Advanced Technology, Indore 452013, India}
\affiliation{Homi Bhabha National Institute, Training School Complex, Anushakti Nagar, Mumbai, 400094, India}
\author{P. Mahadevan}
\affiliation{Department of Condensed Matter and Materials Physics, S. N. Bose National Centre for Basic Sciences, Kolkata, West Bengal, India, 700106.}
\author{S.\ Thirupathaiah}
\email{setti@bose.res.in}
\affiliation{Department of Condensed Matter and Materials Physics, S. N. Bose National Centre for Basic Sciences, Kolkata, West Bengal, India, 700106.}

%

\date{\today}

\begin{abstract}
We present a comprehensive investigation of the structural, electrical transport, and magnetic properties of FeSe$_{\it{x}}$ ($\it{x}$ = 1.14, 1.18, 1.23, 1.28, and 1.32) to unravel the mechanism of the metal-insulator transition observed in these systems. For this, we systematically evaluated the structural parameters of FeSe$_{\it{x}}$ as a function of Se concentration and temperature. We observe increased lattice constants and cell volume with increased Se concentration. On the other hand, the temperature-dependent XRD studies suggest unusual lattice change around the metal-insulator (MI) transition temperature of the respective compositions. This remarkable observation suggests that the anomalous lattice effect originates the MI transition in these systems. Additionally, our density of states (DOS) calculations on FeSe$_{1.14}$  qualitatively explain the MI transition, as the low-temperature (50 K) structure DOS suggests a metallic nature and the high-temperature (300 K) structure DOS shows a gap near the Fermi level.
\end{abstract}

\maketitle
\section{Introduction}
Over the past decade, the iron-chalcogenides have captivated significant research interests due to their remarkable array of physical properties ranging from superconductivity to optical, thermal, and magnetic properties attributed to their diverse structural phases~\cite{hsu2008,hu2021,xu2022}.
Also, the Fe-Se-based systems reveal a range of stable structural phases under ambient conditions, including FeSe$_2$ (orthorhombic), Fe$_3$Se$_4$ (monoclinic NiAs-type), Fe$_7$Se$_8$ (hexagonal NiAs-type), and FeSe (tetragonal PbO type)~\cite{okazaki1956,tewari2020,zhao2022}. The physical properties of these Fe-Se-based systems span across  metals, semiconductors, superconductors, and with varying magnetic structures from paramagnetic to ferrimagnetic, all influenced by the Fe/Se ratio present in the system~\cite{okazaki1956, Liu2012,tewari2020,zhao2022,ghalawat2022}.
These systems can be represented by the chemical formula  FeSe$_{\it{x}}$ ($\it{x} \geq 1$), exhibiting Fe-vacancy in their stable phases. For $\it{x} = 1-1.13$, FeSe$_{\it{x}}$ adopts a PbO-type structure ($\alpha$-phase), while for $\it{x} = 1.14-1.33$, it adopts a NiAs-type structure ($\beta$-phase)~\cite{hagg1933}. Further, a neutron diffraction study on  FeSe$_{1.14}$  demonstrates that the in-plane Fe moments ferromagnetically order while the out-of-plane Fe moments orders antiferromagnetically, resulting in a ferrimagnetic system with a Néel temperature of 455 K,  indicating a great potential for the room temperature spintronics applications~\cite{kawaminami1970}.

 \begin{figure*}[t]
    \centering
    \includegraphics[width=\linewidth]{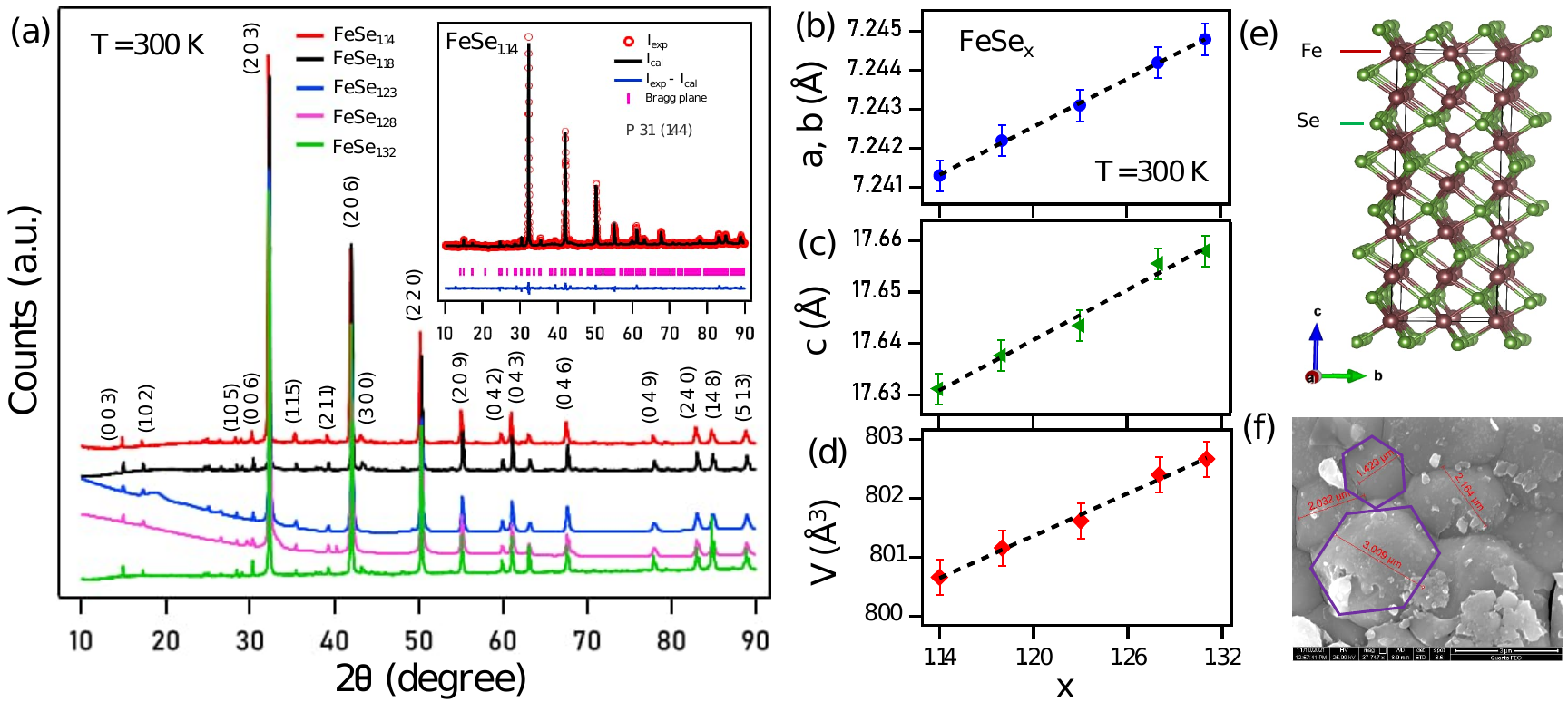}
    \caption{(a) Powder X-ray diffraction (XRD) patterns of FeSe$_{\it{x}}$ ($\it{x}$ = 1.14, 1.18, 1.23, 1.28, and 1.32) measured at room temperature. The inset in (a) shows Rietveld refinement of FeSe$_{1.14}$ XRD data. (b)–(d) Variation of lattice parameters ($a(b), c$) and unit cell volume ($V$) with selenium concentration ($\it{x}$). (e) Trigonal crystal structure of FeSe$_{\it{x}}$. (f) Scanning electron microscopy (SEM) image of the FeSe$_{1.14}$ pellet.}
    \label{1}
\end{figure*}

In these systems, the Fe vacancies are typically ordered in alternating layers, leading to the formation of various superstructures such as $3c$, $4c$, $5c$, and $6c$, depending on the Fe or Se concentration~\cite{andresen1964,kawaminami1967,kawaminami1970}. In addition, FeSe$_{1.14}$ exhibits a spin-reorientation transition at around 100 K, where the easy axis of magnetization shifts from the $ab$-plane to the $c$-axis, causing a significant magnetocaloric effect over a wide temperature range~\cite{adachi1968,radelytskyi2018, Konopelnyk2022}. M$\ddot{o}$ssbauer studies FeSe$_{1.14}$ suggest itinerant Fe $3d$ electrons~\cite{ok1993}, while its nanoscopic form exhibits giant coercivity, indicating a mixed valence state of Fe~\cite{lyubutin2014}. Interestingly, a metal-insulator transition  at around 100 K has been reported earlier in the polycrystalline FeSe$_{1.14}$ from the electrical resistivity measurements that are ascribed to the spin-reorientation transition~\cite{li2016,zhao2022}. On the other hand, the transport measurements on FeSe$_{1.14}$ single crystals having $3c$ superstructure indicate a metallic nature throughout the measured temperature range of 3-300 K without any MI transition~\cite{pajaczkowska1974,ying2016}. Another study demonstrates that the resistivity of FeSe$_{1.14}$ is metallic and temperature dependent below T$_{SR}$, but above T$_{SR}$ it is temperature independent~\cite{kawaminami1967}. Thus, these systems' electrical transport properties, particularly the mechanism of metal-insulator transition, is still elusive. Interestingly,  we prepared FeSe$_{1.14}$ composition that shows the spin-reorientation at around 100 K but the MI transition is found at around 250 K. This is in quite  contrast to the previous reports on MI transition and spin-reorientation transitions found at around 100 K in these systems~\cite{li2016,zhao2022}. This is surprising if one considers the spin-reorientation transition as the mechanism of MI transition as sugegsted earlier~\cite{li2016,zhao2022}. Moreover, the recent studies on polycrystalline Fe$_{1.01-x}$Cu$_x$Se and single crystalline Li$_x$Fe$_7$Se$_8$ suggest lattice disorder-originated MI transition~\cite{PhysRev1958,PhysRevB2010,ying2016}. Thus, the mechanism of metal-insulator transition in FeSe$_{\it{x}}$ systems is yet to be clearly established.

In this manuscript, we aim at unravelling the mechanism of the metal-insulator transition observed in FeSe$_{\it{x}}$ systems.  In order to achieve this, we prepared polycrystalline FeSe$_{\it{x}}$ compositions with various Se concentrations ($\it{x}$ = 1.14, 1.18, 1.23, 1.28, and 1.32)  and performed a comprehensive investigation of the structural, electrical transport, and magnetic properties.  Foremost,  all these systems are found to crystallize in a trigonal crystal structure with a space group of $P3_121$.  The structural analysis reveal that the lattice parameters ($a(b)~and~ c$) show anomalous behaviour around the metal-insulator transition temperature within the temperature range of 185-279 K. Magnetic properties studies exhibit spin-reorientation transition within the temperature range 100-115 K, which is far away from the MI transition temperatures. This remarkable observation suggest an anomalous lattice effect-originated  MI transition in these systems.   To further confirm the structural-originated MI transition, we performed density functional theory (DFT) calculations using the lattice parameters obtained below and above the MI transition temperature.  Our DFT calculations demonstrate reduction in the density of states near the Fermi level for the high-temperature structure (300 K) compared to the low-temperature (50 K) one. The temperature dependent valence band (VB) spectra measured using the photoemission spectroscopy also demonstrate the reduction in spectral density near the Fermi level with increasing temperature.


 \begin{figure*}[t]
    \centering
    \includegraphics[width=\linewidth]{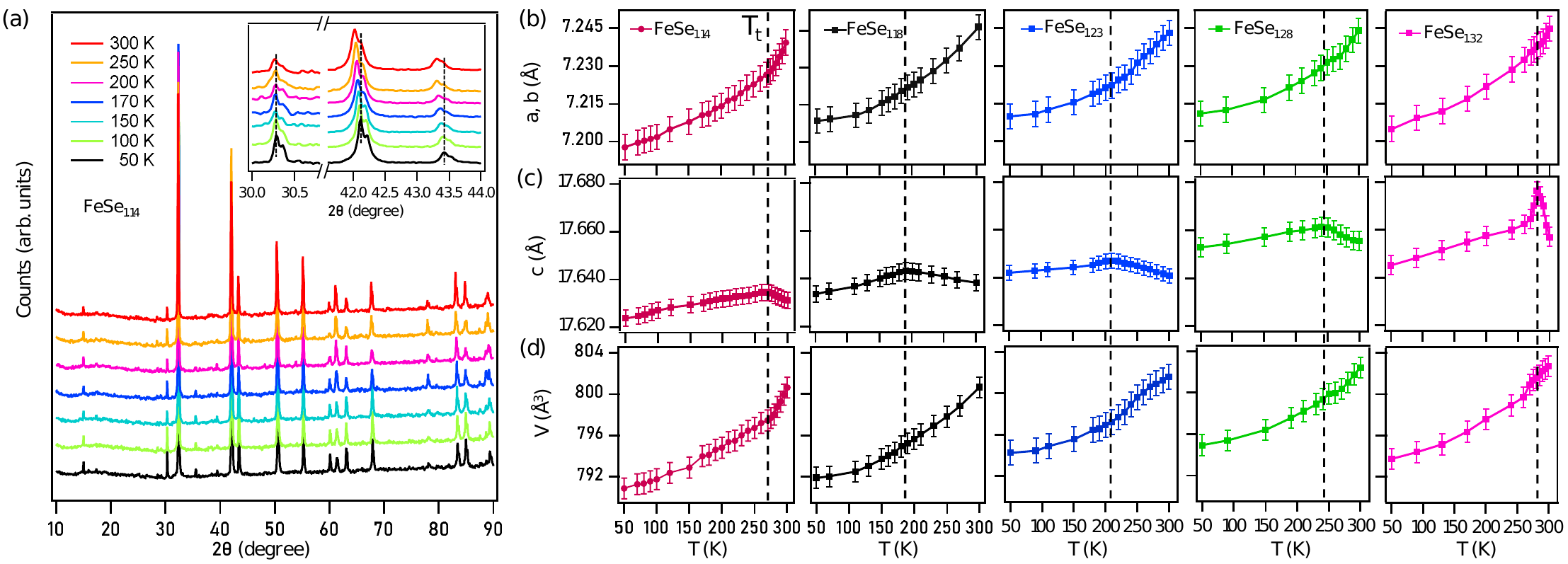}
    \caption{(a) X-ray powder diffraction patterns were collected at different temperatures on FeSe$_{1.14}$. Inset in (a) shows the shift of (0 0 6) and (2 0 6) Bragg's peak positions with temperature. Temperature evolution of lattice parameters  $a(b)$ (b) and $c$ (c). (d) Temperature evolution of unit cell volume ($V$). In the figure, $T_t$ represents the anomalous lattice effect critical temperature for the respective samples.}
    \label{2}
\end{figure*}

\section{Experimental and Theoretical Details}
\subsection{Synthesis and Characterization Techniques}
Polycrystalline FeSe$_{\it{x}}$ with $\it{x}$ = 1.14, 1.18, 1.23, 1.28, and 1.32 were synthesized by the standard solid-state reaction method in evacuated quartz ampoule. In this method, the stoichiometric ratio of Iron (4N, Alfa Aesar) and Selenium powders (4N, Alfa Aesar) were weighed and ground thoroughly in an argon-filled glove box before sealing the mixture in an evacuated quartz ampoule. The quartz ampoule with the powder mixture was slowly heated up to 650 $^{\circ}$C for 10 hours and kept at this temperature for the next three days in a muffle furnace~\cite{andresen1964,akramov2022}. After three days of reaction, the ampoule was slowly cooled to room temperature. As-obtained final compositions were ground thoroughly again in an argon atmosphere, pressed using a two-torr pressure pelletizer, and then annealed at 650 $^{\circ}$C  for two more days. All the prepared samples were stored in an argon-filled glove box to protect them from oxidation.

The phase purity of the compositions was examined by the X-ray diffraction (XRD) technique using a Rigaku X-ray diffractometer (SmartLab, 9kW) equipped with Cu K$_\alpha$ radiation of wavelength 1.5406 $\AA$. The XRD measurements were performed within the temperature range of 3-300 K. Quantitative analysis of chemical compositions was performed using energy dispersive X-ray spectroscopy (EDS), and the morphology of crystallites was examined using the scanning electron microscope (SEM).

Electrical resistivity and magnetic measurements [$M(T)$ and $M(H)$] were carried out on the physical property measurement system (PPMS, 9 Tesla, Quantum Design-DynaCool). The electrical resistivity was measured in a standard four-probe method with an alternating current of 5 mA. Copper leads were attached to the sample using EPO-TEK H21D silver epoxy.

High-resolution X-ray photoelectron spectroscopy (XPS) measurements were conducted for the valence band (VB) spectra at 50 K (LT) and 300 K (HT) in Indus-2 synchrotron radiation centre at ARPES BL-10 beamline,  equipped with PHOIBOS 150 electron analyzer. Atomically clean surface was achieved through argon ion sputtering at 1.5 keV for 1 hour. The valence band (VB) spectra were recorded using a photon energy of 75 eV with the energy resolution set at 30 meV for LT and 100 meV for RT.  The measurement chamber vacuum was maintained at 7$\times$10$^{-11}$ mbar during the measurements.

\subsection{Density Functional Theory}

Electronic structure calculations were performed by density functional theory using plane-wave projected augmented basis set ~\cite{P.E.Blochl1994, G.Kresse1999} and generalized gradient approximation (GGA) exchange-correlation functional ~\cite{JohnP.Perdew1996}  as implemented in the Vienna Ab initio Simulation Package (VASP)~\cite{G.Kresse1996,kresse1996,G.Kresse1993,G.Kresse1994}. In addition, orbital-dependent Coulomb interactions of 4 eV on the Fe $3d$ orbitals were included within the Dudarev implementation~\cite{S.L.Dudarev1998}. The experimental lattice parameters for the low-temperature structure at 50 K and the high-temperature structure at 300 K were used in the calculations, while the atoms were displaced till the forces on them were less than 5 meV/\AA. A gamma-centered $k$-points grid of 6$\times$6$\times$3 was used for the self-consistent calculations and a cutoff energy of 500 eV for the plane waves included in the basis. A $k$-point grid of 8$\times$8$\times$5 points was used for further analysis of the electronic structure.

\section{Results and Discussion}

\subsection{Structural Properties}

Figure~\ref{1}(a) presents the room temperature X-ray diffraction (XRD) patterns for the iron selenide series FeSe$_{\it{x}}$ with varying selenium concentrations ($\it{x}$ = 1.14, 1.18, 1.23, 1.28, and 1.32). The XRD analysis reveals that all diffraction peaks correspond to a pseudo-NiAs-type crystal structure, consistent with the trigonal P3$_1$21 space group (No. 144), as previously reported \cite{harris2019}. Importantly, the absence of secondary phases, such as Fe$_3$Se$_4$ (monoclinic) \cite{andresen1968}, FeSe$_2$ (orthorhombic) \cite{lavina2018}, or FeSe (tetragonal)~\cite{millican2009}, confirms the phase purity of our synthesized compounds.
A notable feature in the diffraction pattern is the (0 0 3) Bragg peak, attributed to iron vacancies within the lattice. These vacancies are situated at specific crystallographic positions, namely (1/2, 1/2, 0; 1/2, 0, 1/3; 0, 1/2, 2/3), which results in the formation of a $3c$ superstructure, as documented in earlier studies~\cite{andresen1964}. The inset of Fig.~\ref{1}(a) shows the Rietveld refinement of the XRD data for FeSe$_{1.14}$ at 300 K, demonstrating an excellent agreement between the experimental data and the fitted model. Fig. S1 of the Supplemental Information details the Rietveld refinements for other compositions~\cite{Supp}.
The derived lattice parameters and unit cell volumes for all compositions are summarized in Tab.~ST1 of the Supplemental Information~\cite{Supp}. These values show a qualitative agreement with previously published data on these systems~\cite{okazaki1956,okazaki1961,andresen1964,li2016,zhao2022}. As shown in Fig.~\ref{1}(b-d), increasing selenium concentration causes a monotonic expansion of the lattice constants and unit cell volume, likely due to additional lattice strain and increased Fe vacancy formation~\cite{PhysRevA.43.3161}. Fig.~\ref{1}(e) illustrates the crystal structure of FeSe$_{1.14}$, featuring 21 Fe atoms and 24 Se atoms per unit cell. Fig.~\ref{1}(f) presents a SEM image of FeSe$_{1.14}$, revealing hexagonal-shaped crystallites with typical sizes ranging from 4 to 9 $\mu m^2$. SEM images for other compositions are provided in Fig.~S2 of the Supplemental Information~\cite{Supp}.

Intriguingly, we have also synthesized FeSe$_{1.38}$, which crystallizes in a monoclinic phase with the $C2/m$ space group, as shown in Fig.~S5(a) of the Supplemental Information~\cite{Supp}. This suggests that the trigonal phase with the $3c$ superstructure is the only stable configuration within the FeSe$_{\it{x}}$ series for 1.14 $\leq$ x $\leq$ 1.32. This finding highlights a critical structural boundary within the FeSe$_{\it{x}}$ system, where the stability of the trigonal phase is preserved up to a specific selenium concentration, beyond which a transition to the monoclinic phase occurs~\cite{tewari2020}.

Figure~\ref{2} presents the temperature-dependent XRD patterns of FeSe$_{1.14}$ measured between 50 and 300 K. All patterns are indexed to the same trigonal crystal symmetry, with Bragg peaks shifting to higher diffraction angles as the temperature increases, indicating the changes in the lattice parameters. The inset of Fig.~\ref{2}(a) highlights this shift. No splitting of Bragg peaks was observed, suggesting the absence of a structural transition within this temperature range. Panels in Figs.~\ref{2}(b) and \ref{2}(c) display the lattice constants $a(b)$ and $c$ as a function of temperature for FeSe$_{\it{x}}$ ($\it{x}=$1.14, 1.18, 1.23, 1.28, 1.32), derived from Rietveld refinement. For FeSe$_{1.14}$, $a(b)$ increases linearly with temperature up to 260 K, after which it rapidly rises, while $c$ increases linearly up to 260 K before decreasing. This indicates an unusual lattice behavior around a critical transition temperature of $T_t \approx 260$ K. Similar behaviors are observed in the other compositions at their respective critical transition temperatures ($T_t\approx$ 180 K, 205 K, 245 K, and 275 K for $\it{x}=$ 1.18, 1.23, 1.28, and 1.32, respectively). Despite these anomalies, the monotonic expansion of unit cell volume ($V$) with temperature, as shown in Fig.~\ref{2}(d), suggests that elongation in the $ab$-plane dominates over contraction in the $c$ direction~\cite{radelytskyi2018}. Temperature-dependent XRD data for other Se concentrations are provided in Fig.~S3 of Supplemental Information~\cite{Supp}.

\begin{figure}[t]
    \centering
    \includegraphics[width=\linewidth]{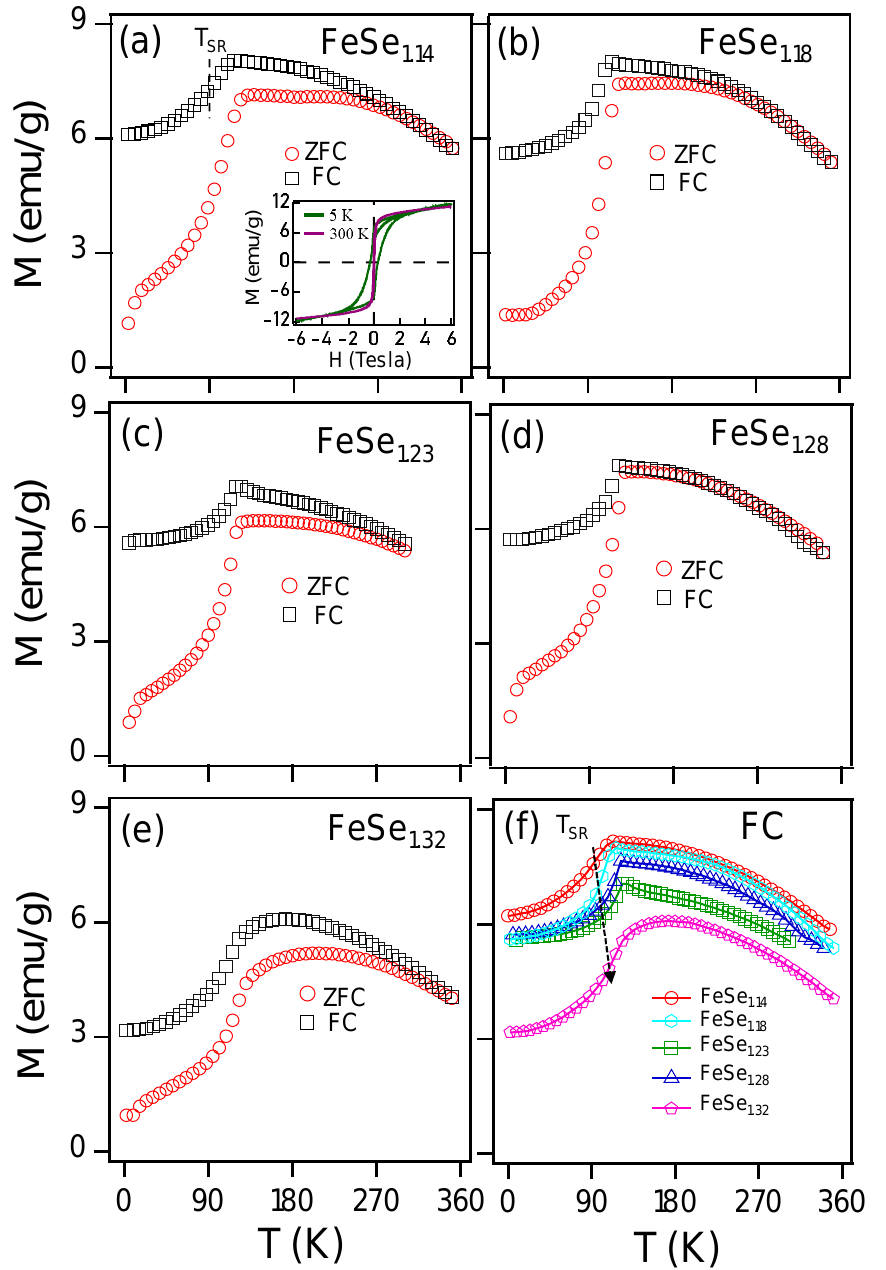}
    \caption{(a)–(e) Temperature-dependent magnetization, $M(T)$, measured in zero-field cooled (ZFC) and field-cooled (FC) modes with an applied field of 1000 Oe for FeSe$_{\it{x}}$ ($\it{x}$ = 1.14, 1.18, 1.23, 1.28, and 1.32). The inset in (a) displays the magnetization as a function of field, $M(H)$, at T = 5 K and 300 K. In (a), $T_{SR}$ represents the spin-reorientation transition, determined by $dM/dT$. (f) Comparison of temperature-dependent magnetization of FeSe$_{\it{x}}$ compositions, demonstrating the evolution of $T_{SR}$ with increasing $\it{x}$.}
    \label{3}
\end{figure}

The anomalous lattice behaviour with temperature, as previously observed in other systems,  is likely happening due to local structure distortions~\cite{okazaki1961, louca1997,petkov2022}. In iron chalcogenides,  FeX$_{1+y}$ ($X$ = S, Se),  the Fe cations octahedrally coordinates with six $X$ anions to form FeX$_6$ octahedra. Whereas the $X$ anions reside within trigonal prisms created by six Fe atoms, known as XFe$_6$ prisms~\cite{wold-1993,lyubutin2014}. This arrangement harbors two distinct Fe ionic states (Fe$^{2+}$ and Fe$^{3+}$),  along with the Fe vacancies, leading complex structural behavior~\cite{andresen1964,andresen1968}. In addition, the Fe $3d$-orbitals split into threefold degenerate $t_{2g}$ and twofold degenerate $e_g$ states under the influence of crystal field splitting. The $t_{2g}$ and $e_g$ degeneracy are further lifted by the Jahn-Teller distortion, leading to additional symmetry breaking and new orbital energy levels~\cite{ghalawat2022}. This kind of structural reconfiguration is pivotal in influencing the material's physical properties. Furthermore, changes in the Fe-Fe bond length with temperature, as observed by I. Radelytskyi $et~al.$ in FeSe$_{1.14}$ (Fe$_7$Se$_8$), underscores the significance of temperature-dependent structural dynamics in these systems~\cite{radelytskyi2018}.

\subsection{Magnetic Properties}

Temperature-dependent magnetization ($M(T)$) of FeSe$_{\it{x}}$ ($\it{x}$ = 1.14, 1.18, 1.23, 1.28, and 1.32) is shown in Figs.~\ref{3}(a)-~\ref{3}(e),  measured with the zero-field-cooled (ZFC) and field-cooled (FC) modes under an applied magnetic field of 1000 Oe. Consistent with previous reports on FeSe$_{1.14}$, the spin-reorientation transition from the easy $c$-axis to easy $ab$-plane has been observed at $T_{SR}$ $\approx$ 100 K [see Fig.~\ref{3}(a)], leading to an abrupt drop in the magnetization below $T_{SR}$~\cite{radelytskyi2018}. The T$_{SR}$ $\approx$ 100 K also confirms the $3c$ superstructure present in the studied system~\cite{kawaminami1967,kawaminami1970,li2016,zhao2022} as $T_{SR}$ $\approx$ 220 K is recorded for the  $4c$ superstructure~\cite{terzieff1978}. A similar spin-reorientation transition has been observed from the other Se concentration samples at their respective transition temperatures. Fig.~\ref{3}(f) shows slight change in the spin-reorientation temperature with increasing Se concentration from $T_{SR}$ $\approx$ 100 K for FeSe$_{1.14}$ to $T_{SR}$ $\approx$ 115 K for FeSe$_{1.32}$. Further, the non-overlapping ZFC and FC curves and the magnetic hysteresis observed at 5 K suggest complex magnetic interactions in FeSe$_{\it{x}}$. This complexity arises from the coexistence of Fe$^{2+}$ and Fe$^{3+}$ states with ordered Fe vacancies~\cite{andresen1964,lin2009}, leading to exchange, superexchange, and double exchange interactions~\cite{Zhang2019}.

To further investigate the magnetic state of FeSe$_{\it{x}}$, we plotted magnetization as a function of applied field [$M(H)$] in the inset of Fig.~\ref{3}(a) for FeSe$_{1.14}$ measured at 5 and 300 K. The observed saturation magnetization ($M_s$) of 2.32 $\mu_B$/f.u. at 5 K aligns well with the previously reported values from similar systems~\cite{li2016}. $M(H)$ data for FeSe$_{\it{x}}$ ($\it{x}$ = 1.18, 1.23, 1.28, and 1.32) are shown in Fig. S4 of the Supplemental Information~\cite{Supp}. The magnetic hysteresis loop at 5 K exhibits a high coercive field (H$_c$) of 3000 Oe, consistent with previous observations in FeSe$_x$~\cite{lyubutin2014,zhao2022}, and the magnetization remains unsaturated up to 6 T, further confirming its ferrimagnetic nature at 5 K~\cite{ghalawat2022,gao2011,akramov2022}. But at 300 K,  despite the overall magnetic nature of the system is still the ferromagnetic, the magnetic hysteresis disappears.


\begin{figure}[t]
    \centering
    \includegraphics[width=\linewidth]{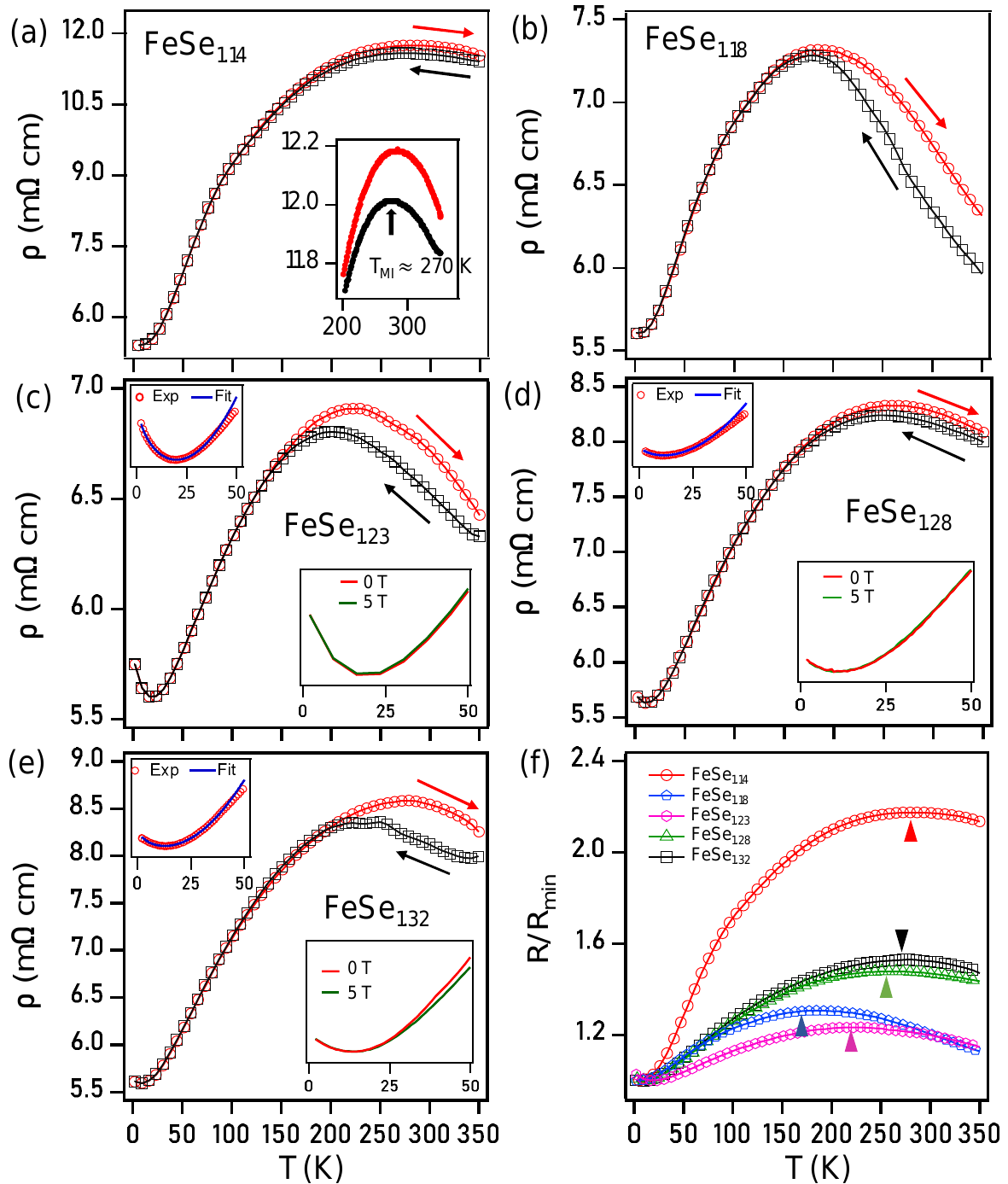}
    \caption{(a)–(e) Temperature-dependent longitudinal electrical resistivity, $\rho(T)$, for polycrystalline FeSe$_{\it{x}}$ ($\it{x}$ = 1.14, 1.18, 1.23, 1.28, and 1.32) measured from 2 to 350 K during both heating and cooling cycles. Inset in (a) demonstrates metal-insulator transition ($T_{MI}$) for FeSe$_{1.14}$. The top insets of (c), (d), and (e) show the low-temperature data fitted using the Eqn.~\ref{Eq1}. The bottom insets of (c), (d), and (e) display $\rho(T)$ under applied magnetic fields of H = 0 and 5 T. Panel (f) presents the normalized resistance measured during the heating cycle for FeSe$_{\it{x}}$. In (f), the arrows point the $T_{MI}$ of various FeSe$_{\it{x}}$ compositions.}
    \label{4}
\end{figure}

\begin{figure}[b]
    \centering
    \includegraphics[width=\linewidth]{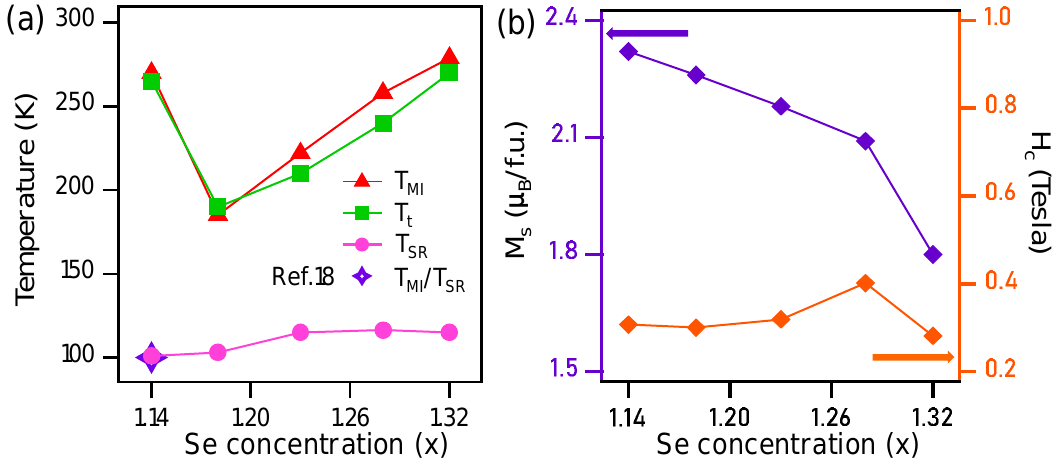}
    \caption{(a) Spin-reorientation transition (T$_{SR}$),  metal-insulator transition (T$_{MI}$), and anomalous lattice change ($T_t$) temperature are plotted as a function of Se concentration ($\it{x}$). (b) The saturation magnetization ($H_s$) and magnetic coercivity ($H_c$) are plotted as a function of Se concentration ($\it{x}$). Plots shown in this figure are summarized as tables in Tab. S2 and Tab. S3 of the Supplemental Information~\cite{Supp}.}
    \label{5}
\end{figure}

\subsection{Electrical Transport Properties}

Figures~\ref{4}(a)-~\ref{4}(e) present the longitudinal electrical resistivity, $\rho(T)$, measured within the temperature range of 2 and 350 K for FeSe$_{\it{x}}$. The zoomed-in image shown in the inset of Fig.~\ref{4}(a) demonstrates the metal-insulator transition temperature ($T_{MI}$) at approximately 270 K for FeSe$_{1.14}$. Similar metal-insulator transitions were observed from the other compositions as well at approximately 179 K, 198 K, 244 K, and 279 K, respectively, for $\it{x}$ = 1.18, 1.23, 1.28, and 1.32, as depicted in Fig.~\ref{4}(f). Additionally, we detected thermal hysteresis near T$_{MI}$ during the heating and cooling cycles of the resistivity measurements [see Figs.~\ref{4}(a)-(e)], underlining   the structural changes around these temperatures. This is in agreement with the anomalous lattice behaviour observed in Figs.~\ref{2}(b) and ~\ref{2}(c).

\begin{figure*}[t]
    \centering
    \includegraphics[width=\linewidth]{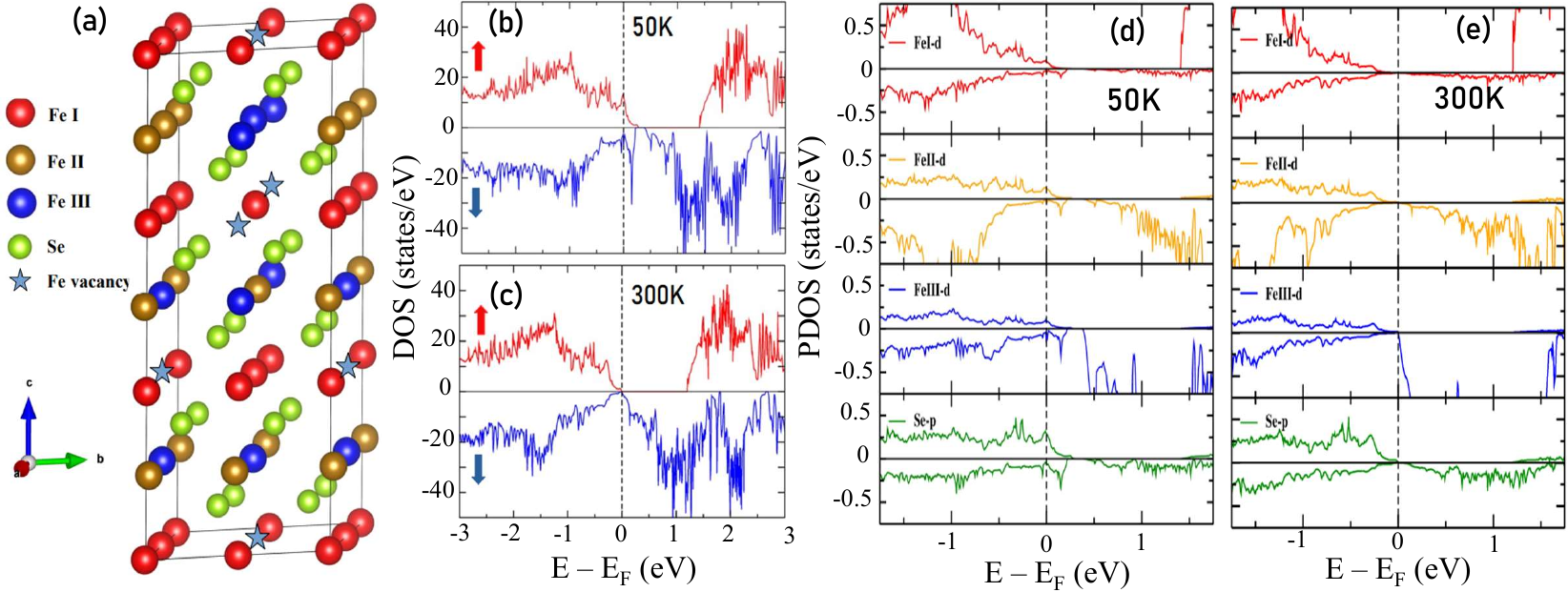}
    \caption{The schematic presentation of the crystal structure of FeSe$_{1.14}$ illustrates the positions of various types of Fe (FeI, FeII, and FeIII) atoms and Fe vacancies. The total density of states (DOS) at 50 K (b) and 300 K (c) is shown with the red and blue lines representing the up-spin and down-spin channels, respectively. The projected DOS of the Fe $3d$ orbital contribution of each type of Fe (FeI, FeII, and FeIII) atoms and the Se $4p$ orbital contribution near Fe(II) are depicted at 50 K (d) and 300 K (e).}
    \label{6}
\end{figure*}

The metal-insulator transition, as observed in the resistivity data, can stem from various reasons, including the electron correlations~\cite{Kiryukhin1997, Lee2001}, structural phase transitions~\cite{Dho2001}, or magnetic ordering~\cite{Kiryukhin1997, Imada1998, Dho2001, Sanchez2000}. A subtle alteration in bond lengths, possibly driven by the Jahn-Teller effect, is also known to trigger the MI transitions~\cite{Garcia-Munoz1992, Kiryukhin1997}. While changes in the crystal structure or the emergence of magnetic ordering, such as antiferromagnetism, can impact the electronic band structure, our observation of the decoupled spin-reorientation transitions eliminate the magnetism as the primary cause of the metal-insulator transition in FeSe$_{\it{x}}$ systems. Instead, the anomalous behavior of temperature-dependent lattice parameters [$a(b)~ and~ c$] around $T_{t}$, coupled with significant thermal hysteresis near this transition, as depicted in Figs.~\ref{2} (b) and ~\ref{4}, strongly suggests that the structural distortions are the driving force of MI transition.

In addition to the $T_{MI}$ transition, we observe an upturn in the electrical resistivity with a resistivity minima ($T_m$) at 20 K, 12 K, and 8.5 K for FeSe$_{1.23}$, FeSe$_{1.28}$, and FeSe$_{1.32}$, respectively. Such an upturn in the resistivity can arise from several reasons, including the Kondo effect~\cite{PhysRevB2016}, weak localization~\cite{PhysRevB1980,PhysRevB2016}, or the elastic electron-electron ($e-e$) scattering~\cite{guo2012}. The Kondo effect typically occurs in nonmagnetic systems with diluted magnetic impurities~\cite{barua2017}, but given that all our samples are ordered, we can rule out this possibility. On the other hand, we perfectly fitted the upturn resistivity data using the inelastic $e-e$ scattering  Eqn.~\ref{Eq1}~\cite{guo2012} as illustrated in the top inset of Figs.~\ref{4}(c), ~\ref{4}(d), and ~\ref{4}(e). Thus, the upturn in the resistivity is predominantly driven by the elastic $e-e$ interactions, highlighting the electronic correlations in these systems. Fig.~\ref{4}(f) shows normalized resistance as a function of temperature, demonstrating different $T_{MI}$ for different Se concentrations.

\begin{equation}
   \rho(T) = \rho_0 -\alpha*T^{0.5} +\beta*T^2
   \label{Eq1}
\end{equation}

where $\alpha$ and $\beta$ are the elastic and inelastic $e-e$ scattering coefficients, respectively.

Next, Fig.~\ref{5}(a) shows the spin-reorientation transition temperature ($T_{SR}$) plotted as a function of Se concentration ($\it{x}$). From Fig.~\ref{5}(a), it is clear that $T_{SR}$ slightly increases with increasing $\it{x}$, possibly due to the increase in the indirect exchange coupling strength~\cite{ghalawat2022}. On the other hand, T$_{MI}$  shows a non-monotonic behaviour with increasing $\it{x}$ and matches well with $T_t$ (structural) transition, confirming that the metal-insulator transition is independent of magnetism~\cite{li2016}. Furthermore, as shown in Fig.~\ref{5}(b),  the saturation magnetization [M$_{s}$] extracted from Fig.~\ref{3} and Fig.~S4 drops monotonically with increasing Se concentration, suggesting the robustness of direct exchange interaction (Fe-Fe) in these systems which decreases with increasing Se concentration. Remarkably, we also observe high coercivity H$_{c}$ in these compounds at T = 2 K [see Fig.~\ref{5}(b)] that is almost independent of Se concentration.

\section{Electronic Band Structure}

We examined the temperature-dependent changes in the electronic structure of FeSe$_{1.14}$ by calculating the density of states for the 50 K and 300 K using the lattice parameters determined at these respective temperatures. Our calculations confirm that the ferrimagnetic state has the lowest ground state energy in which the in-plane spins align in parallel. At the same time, they are oriented antiparallel in the out-of-plane direction, considering the layer stacking in the $c$-direction. We observe three types of magnetic moments of Fe in the system on the basis of  their positions. The Fe atoms in the layers that contain the Fe vacancies are denoted as FeI and are situated between vacancy-free Fe layers in which the atoms are classified as FeII (far from the vacancy) and FeIII (near to the vacancy) as shown in Fig.~\ref{6}(a). The magnetic moments on Fe are found to be 3.37 $\mu_B$ for FeI, 3.52  $\mu_B$ for FeII, and 3.76 $\mu_B$ for FeIII from the high-temperature structure, while for the low-temperature structure they are 3.37, 3.49 and 3.76 $\mu_B$, respectively. Hence, the structural changes lead to a small change in the magnetic moment of the FeII atom. However, the effect of structural change is found to be significant on the electronic band structure.

Fig.~\ref{6}(b) shows the spin-polarized density of states calculated for the 50 K temperature structure, while Fig.~\ref{6}(c) shows the same but calculated for the 300 K structure. Specifically, the low-temperature structure is found to be metallic, while a gap opens up in the density of states associated with the high-temperature structure as shown in Figs.~\ref{6} (b) and ~\ref{6}(c). To understand the changes in the electronic structure due to structural modifications, we examined the projected partial density of states for the FeI, FeII, FeIII 3d, and Se 4p states shown in Figs.~\ref{6}(d) and ~\ref{6}(e). In the partial density of states, we observe that Fe has almost fully occupied up-spin and partially occupied down-spin states in each case. As the lattice parameter ($a$, $b$) of the structure increases with temperature, an overall increase in Fe-Se bond length in going from the low-temperature structure to the high-temperature structure is observed. An increase in the Fe-Se bond length decreases the Fe-Se hopping interaction strength, pulling the anti-bonding states deeper into the valence band and thus driving the system to an insulating state at high temperatures.

\begin{figure*}
    \centering
    \includegraphics[width=0.75\linewidth]{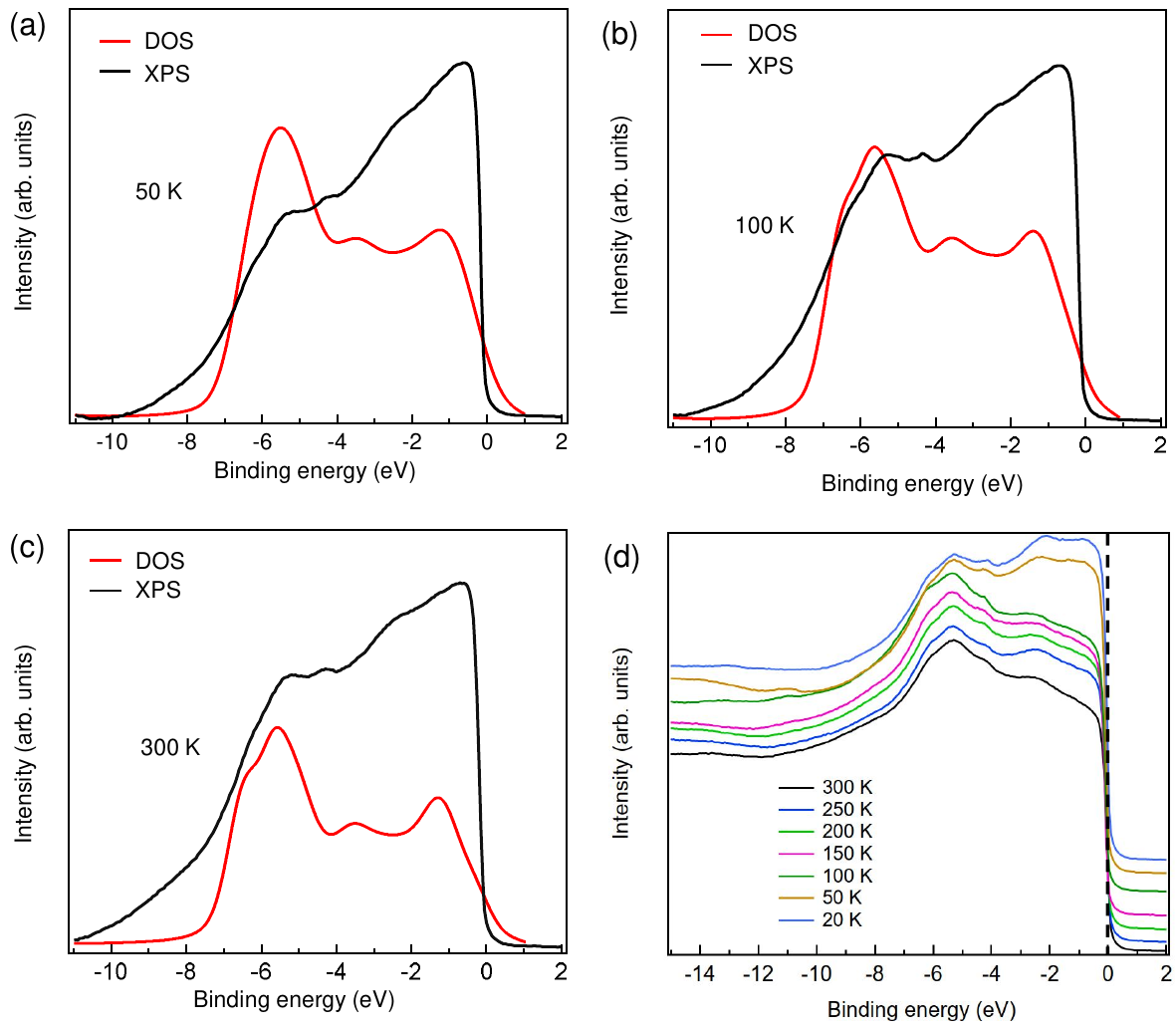}
    \caption{X-ray photoelectron spectroscopy (XPS) measurements on FeSe$_{1.14}$. (a)-(c) Calculated density of states (DOS) overlapped with valence band (VB) spectra obtained using XPS.   (d) Temperature-dependent valence band spectra. The data in (d) is offset vertically for a better visualization.  In figure, the VB spectra is measured with a photon energy of 75 eV.}
    \label{7}
\end{figure*}

We further measured the valence band spectra using the X-ray photoelectron spectroscopy (XPS) using 75 eV photon energy on FeSe$_{1.14}$ at various temperatures as shown in Fig.~\ref{7}. In Fig.~\ref{7}, we also compared the experimental spectral features measured at 50, 100 and 300 K with the calculated density of states (DOS) for the structures at respective temperatures. Although the photoemission cross-section of Fe $3d$ is approximately 47 times greater than that of Se $4p$ at a photon energy of 75 eV~\cite{Yeh1985}, the peak positions in both the XPS and DOS data show good agreement. Thus, the experimental valence band spectra support the theoretical prediction of changes in the density of states between the high and low-temperature structures. In addition, from the temperature dependent valence band spectra [see Fig.~\ref{7}(d)], we notice that the spectral intensity near the Fermi level decreases drastically with increasing temperature when compared to the spectral intensity below  Fermi level ($E$=-2.5 eV and -5.7 eV), further supporting the metal-insulator transition observed at the high temperatures as the states near the Fermi level contribute to the electrical transport.  Finally, before the conclusion, we would like highlight the electrical transport and magnetization data of monoclinic FeSe$_{1.38}$ as shown in Fig. S5 of the Supplemental Information~\cite{Supp}. Interestingly, In Fig. S5(b),  we do not find any MI transition in the resistivity data between 2 and 350 K despite the magnetization data clearly shows the spin-reorientation transition at around 112 K. This further supports our argument on the non-magnetic originated MI transition in these systems.

\section{Summary}

In summary, we investigated the metal-insulator (MI) transition in FeSe$_{\it{x}}$ by synthesizing the polycrystalline samples with varying selenium concentrations ($\it{x}$ = 1.14, 1.18, 1.23, 1.28, and 1.32) and performing a comprehensive analysis on their structural, electrical, and magnetic properties. All samples exhibit a trigonal crystal structure with the P3$_1$21 space group.  Magnetization studies reveal spin-reorientation transition, varying between 100-116 K,  that is decoupled with the MI transition, varying between 185 and 279 K. On the other hand, the temperature-dependent XRD studies suggest unusual lattice changes around the metal-insulator (MI) transition temperature of the respective compositions, $T_t$ varying between 190 and 280 K. This remarkable observation suggests that the anomalous lattice effect originates the MI transition in these systems. Additionally, our density of states (DOS) calculations on FeSe$_{1.14}$ at 50 K and 300 K qualitatively explain the MI transition, as the low-temperature (50 K) structure DOS suggests a metallic nature and the high-temperature (300 K) structure DOS shows a gap near the Fermi level.

\section{acknowledgement}
This research has used the Technical Research Centre (TRC) Instrument Facilities of S. N. Bose National Centre for Basic Sciences, established under the TRC project of the Department of Science and Technology. P.M. acknowledges the support from SERB through Project No. SERBPOWER(SPF/2021/000066). The support and the resources provided by Param Yukti Facility under the National Supercomputing Mission, Government of India at the Jawaharlal Nehru Centre For Advanced Scientific Research, Bengaluru are gratefully acknowledged. S.Paul acknowledges the PhD fellowship from DST-INSPIRE(IF190524), Government of India.

\bibliography{FeSe_PRB}
\section*{Supplementary Information}

 \begin{figure*}[h]
    \centering
    \includegraphics[width=\linewidth]{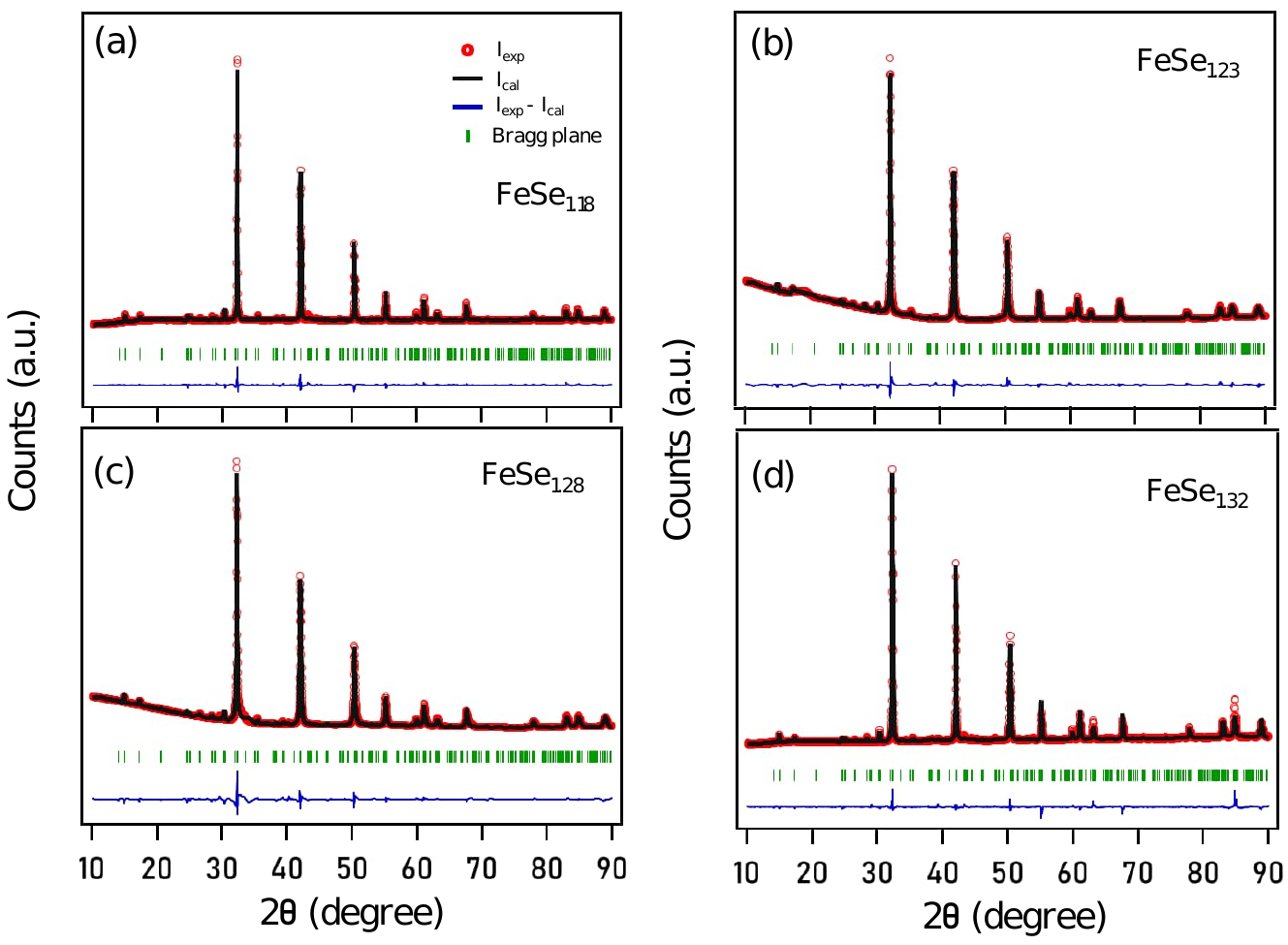}
    \caption{Fig. S1: Room temperature X-ray diffraction patterns of FeSe$_{x}$ with $x = 1.18$, $1.23$, $1.28$, and $1.32$. Rietveld refinement confirms that all diffraction peaks correspond to the trigonal phase with no detectable impurity phases.}
 \label{S1}
\end{figure*}

 \begin{table*}[b!]
\caption{Tab S1: Summary of the structural parameters of FeSe$_{x}$}

\hspace{1cm}
\begin{tabular*}{\linewidth}{c @{\extracolsep{\fill}} ccccc}
 \hline
 Composition & a, b (\AA) & c (\AA) & V (\AA$^3$) & $\chi^2$  \\ [1.5ex]
 \hline
FeSe$_{1.14}$ & 7.2413(3) & 17.6312(4) & 800.661(2) & 2.64 \\ [1.5ex]
 \hline
FeSe$_{1.18}$ & 7.2422(4) & 17.6377(6) & 801.158(4) & 3.01  \\[1.5ex]
 \hline
FeSe$_{1.23}$ & 7.2431(3) & 17.6435(3) & 801.62(4) & 3.44  \\[1.5ex]
 \hline
FeSe$_{1.28}$ & 7.2442(5) & 17.6555(4) & 802.402(2) & 4.53  \\[1.5ex]
 \hline
FeSe$_{1.32}$ & 7.2448(4) & 17.6580(3) & 802.665(4) & 5.03  \\ [1.5ex]
 \hline
\end{tabular*}
\label{ST1}
\end{table*}

\begin{figure*}
    \centering
    \includegraphics[width=\linewidth]{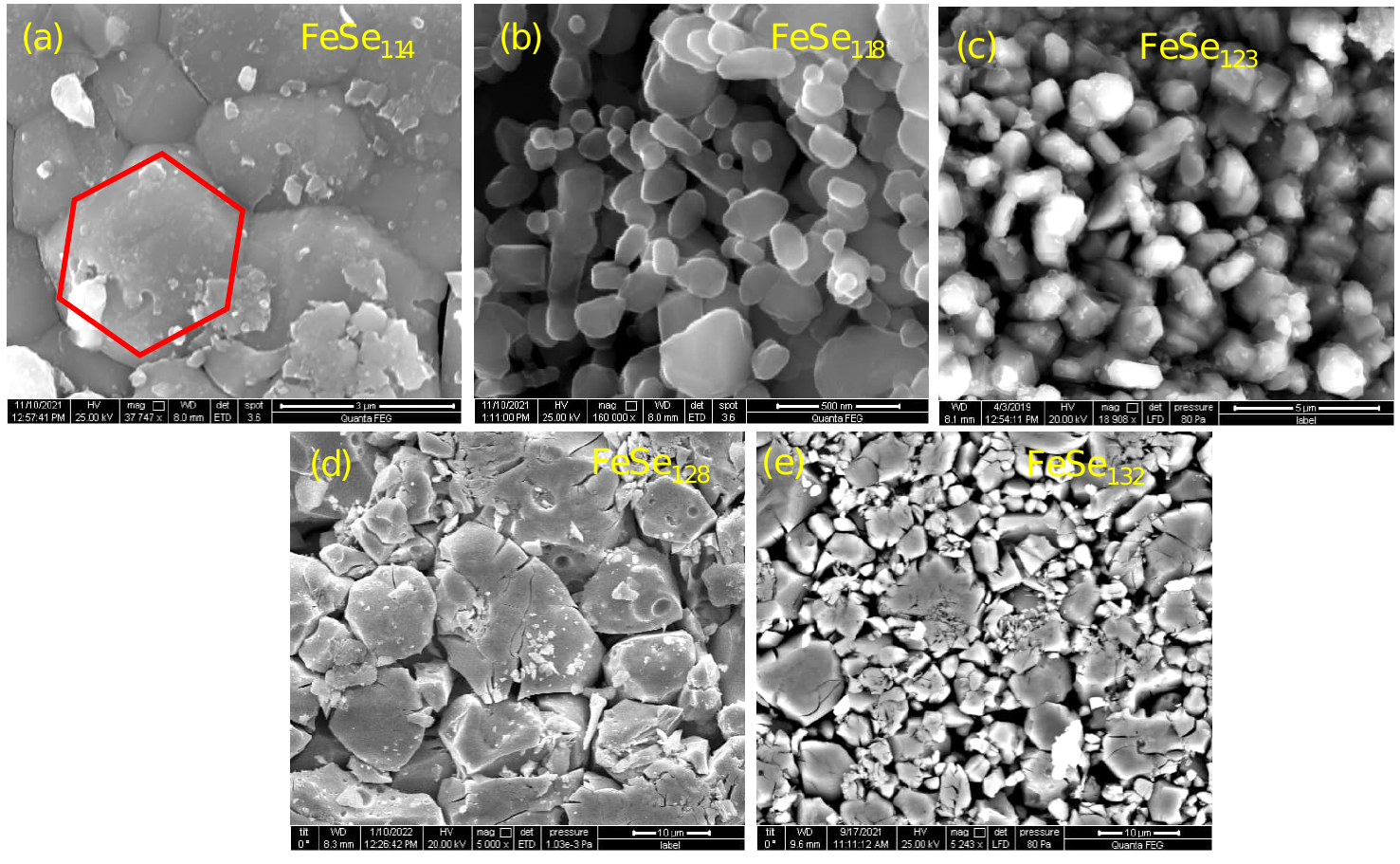}
    \caption{Fig. S2:Field Emission Scanning Electron Microscopy (FE-SEM) images of trigonal FeSe$_{x}$ with $x = 1.14$, $1.18$, $1.23$, $1.28$, and $1.32$.}
    \label{1}
\end{figure*}

\begin{figure*}[h]
    \centering
    \includegraphics[width=\linewidth]{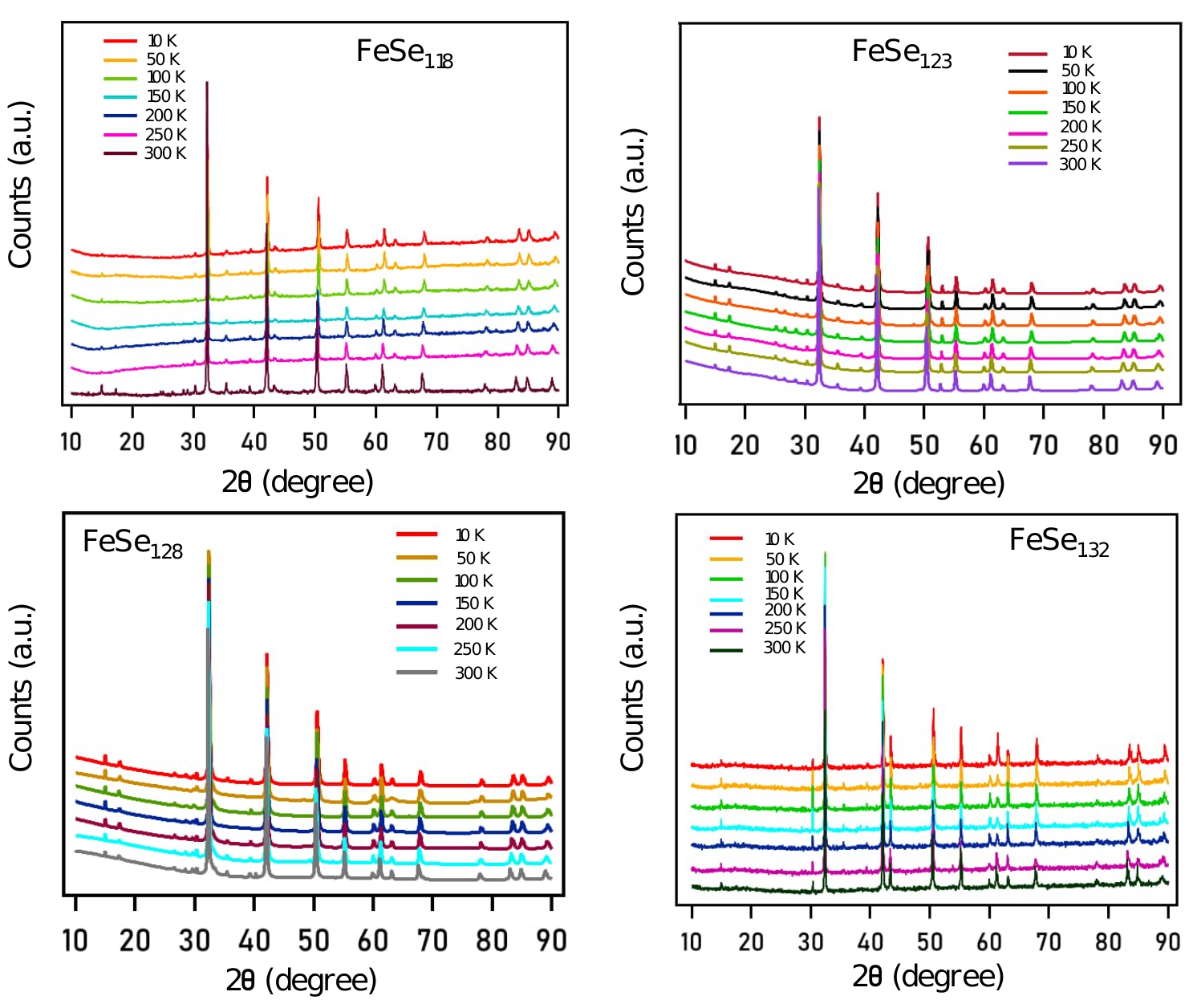}
    \caption{Fig. S3:Temperature-dependent X-ray diffraction (XRD) patterns of FeSe$_{x}$ for $x = 1.18$, $1.23$, $1.28$, and $1.32$, recorded across the temperature range of 10–300 K.}
    \label{S2}
\end{figure*}



\begin{table*}[hbt!]
\caption{Tab S2: Summary of metal-insulator transition (T$_{MI}$), Spin-reorientation transition ($T_{SR}$), anomalous lattice effect temperature ($T_t$), and low-temperature minima due to Coulombic interactions (T$_{CI}$).}
\begin{tabular*}{\linewidth}{c @{\extracolsep{\fill}} ccccc}
 \hline
 Composition & T$_{MI}$ (K) & T$_{SR}$ (K) & T$_{t}$ (K) & T$_{CI}$ (K) \\ [1.5ex]
 \hline\hline
FeSe$_{1.14}$ & 270 & 101 & 265 &  \\ [1.5ex]
 \hline
FeSe$_{1.18}$ & 185 & 103 & 190 &  \\[1.5ex]
 \hline
FeSe$_{1.23}$ & 222 & 115 & 210 & 20 \\[1.5ex]
 \hline
FeSe$_{1.28}$ & 258 & 116 & 240 & 12  \\[1.5ex]
 \hline
FeSe$_{1.32}$ & 279 & 115 & 280 & 8.5 \\ [1.5ex]
 \hline
\end{tabular*}
\label{T2}
\end{table*}

\begin{figure*}
    \centering
    \includegraphics[width=\linewidth]{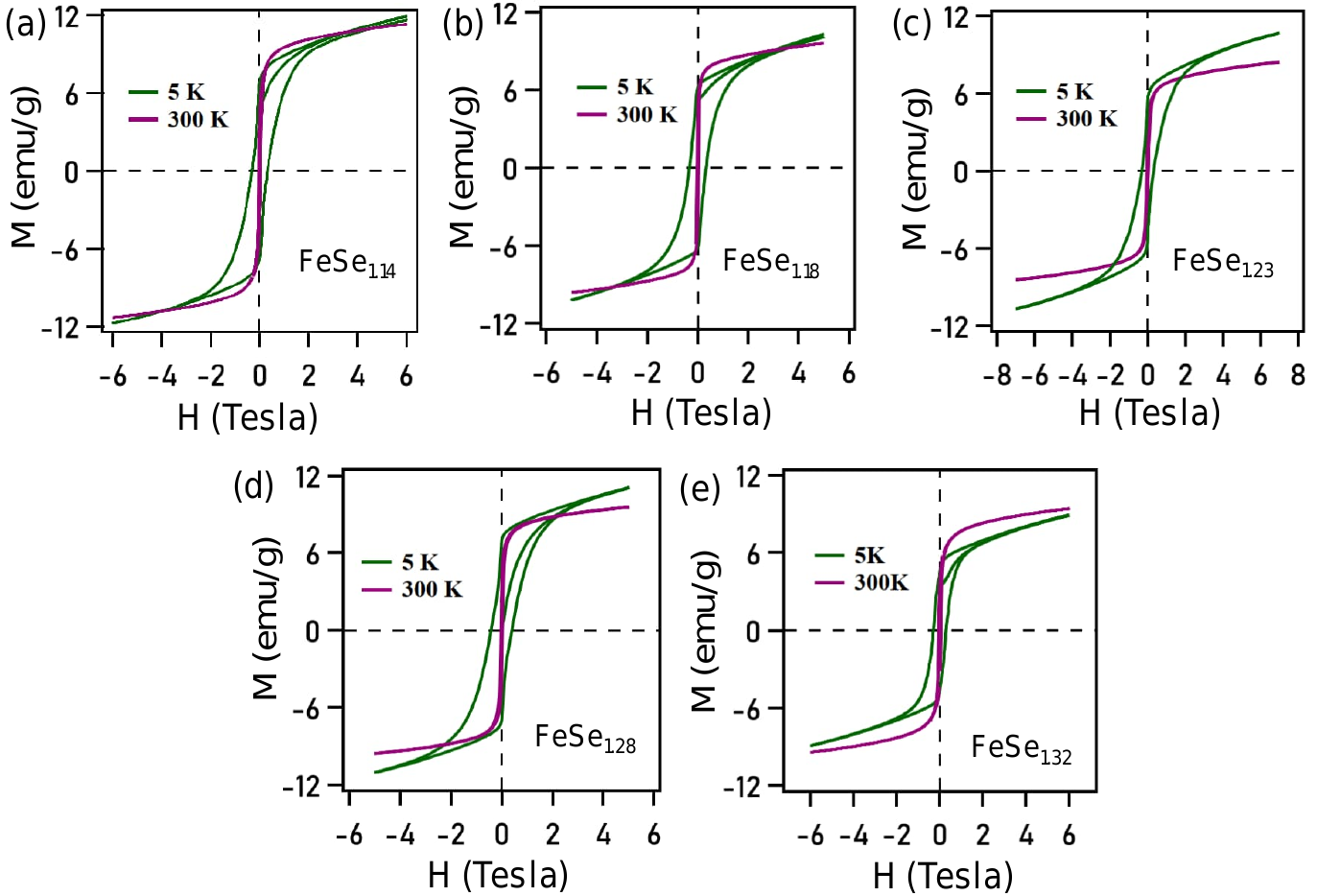}
    \caption{Fig. S4:Magnetization as a function of field [$M(H)$] recorded at 5 and 300 K for FeSe$_{x}$ ($x = 1.14$, $1.18$, $1.23$, $1.28$, and $1.32$.}
    \label{1}
\end{figure*}

\begin{figure*}
    \centering
    \includegraphics[width=\linewidth]{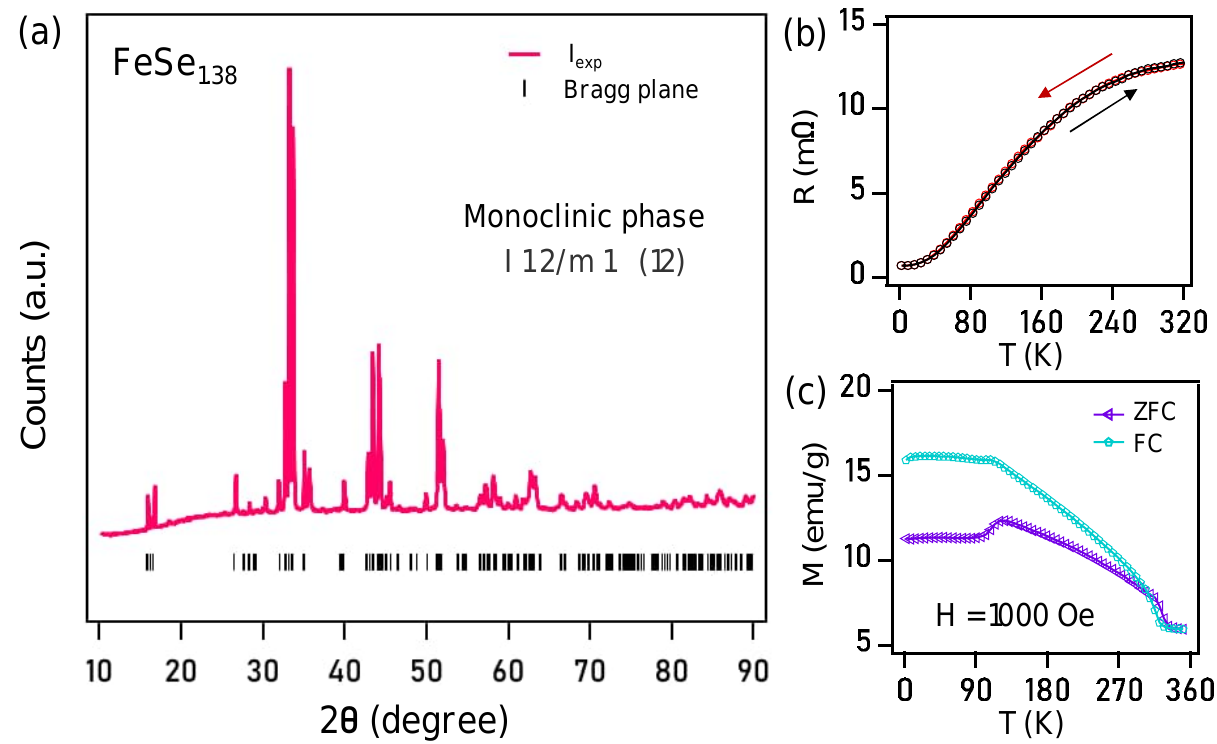}
    \caption{Fig. S5:(a) Powder X-ray diffraction pattern taken at room temperature on FeSe$_{1.38}$.  (b) Temperature-dependent resistance measured during both heating and cooling cycles (absence of hysteresis in the resistance data suggest absence of structural changes in FeSe$_{1.38}$ between 2 and 350 K).  (c) Magnetization as a function of temperature under zero-field-cooling (ZFC) and field-cooling (FC) modes with an applied field of 1000 Oe. In (c), the spin-reorientation transition is found at around 112 K.}
    \label{1}
\end{figure*}

\begin{table*}[t]
\caption{Tab S3: Summary of saturation magnetization ($M_s$) and coercivity ($H_c$) at 5 K.}
\hspace{1cm}
\begin{tabular*}{\linewidth}{c @{\extracolsep{\fill}} ccccc}
 \hline
 Composition &  H$_c$ (Oe) &  M$_s$ ($\mu_B$/f.u.) \\ [1.5ex]
 \hline\hline
FeSe$_{1.14}$ & 3071.6 & 2.32 \\ [1.5ex]
 \hline
FeSe$_{1.18}$ & 3000 & 2.26  \\[1.5ex]
 \hline
FeSe$_{1.23}$ & 3188.3 & 2.18  \\[1.5ex]
 \hline
FeSe$_{1.28}$ & 4015 & 2.09   \\[1.5ex]
 \hline
FeSe$_{1.32}$ & 2813.2 & 1.8  \\ [1.5ex]
 \hline
\end{tabular*}
\label{T2}
\end{table*}


\end{document}